\documentclass[aps,prx,reprint,amsmath,amssymb,longbibliography,floatfix]{revtex4-2}
\usepackage[T1]{fontenc}
\usepackage{lmodern}
\usepackage{bm}
\usepackage{microtype}
\usepackage{xcolor}
\usepackage{tikz}
\usetikzlibrary{arrows.meta}
\definecolor{CommentLink}{rgb}{0,0,0.45}
\usepackage{hyperref}
\hypersetup{colorlinks=true,linkcolor=CommentLink,citecolor=CommentLink,
  urlcolor=CommentLink,
  pdftitle={Comment on Gyrators for superconducting circuit design},
  pdfauthor={Adrian Parra-Rodriguez and I. L. Egusquiza}}
\newcommand{\dd}{\mathrm{d}}
\newcommand{\R}{\mathbb{R}}

\begin{document}
\raggedbottom
\postdisplaypenalty=1000
\title{Comment on ``Gyrators for superconducting circuit design''}
\author{Adrian Parra-Rodriguez}
\email{adrian.parra.rodriguez@gmail.com}
\affiliation{Walther-Meißner-Institut, Bayerische Akademie der Wissenschaften, 85748 Garching, Germany}
\affiliation{Munich Center for Quantum Science and Technology (MCQST), 80799 Munich, Germany}
\affiliation{Technical University of Munich, TUM School of Natural Sciences, Physics Department, 85748 Garching, Germany}
\author{I. L. Egusquiza}
\email{inigo.egusquiza@ehu.es}
\affiliation{Department of Physics, University of the Basque Country UPV/EHU, Apartado 644, 48080 Bilbao, Spain}
\affiliation{EHU Quantum Centre, University of the Basque Country UPV/EHU, Apartado 644, 48080 Bilbao, Spain}

\begin{abstract}
We challenge the novelty claimed in ``Gyrators for superconducting circuit design'', by Salcedo et al., Ref.~\cite{Salcedo:2026}, for its first-order flux--charge formalism, constraint classification, and nonlinear gyrator substitutions, and also challenge its characterization of earlier methods. The underlying construction was established in an uncited  reference~\cite{ParraRodriguez:2024} and
extended to infinite-dimensional circuits in the cited Ref.~\cite{ParraRodriguez:2025}.  The kinetic term proposed in Ref.~\cite{Salcedo:2026}  to account for gyrators is locally a total derivative on the gyrator constraint manifold, so it is subsumed in~\cite{ParraRodriguez:2024}, which imposes Kirchhoff's laws and the constraints of general ideal nonreciprocal multiports jointly at the outset. The contributions of Ref.~\cite{Salcedo:2026} reduce to particular reciprocal realizations and software.  We analyze critically Ref.~\cite{Salcedo:2026} and provide explicit contrast with previous literature.
\end{abstract}
\maketitle

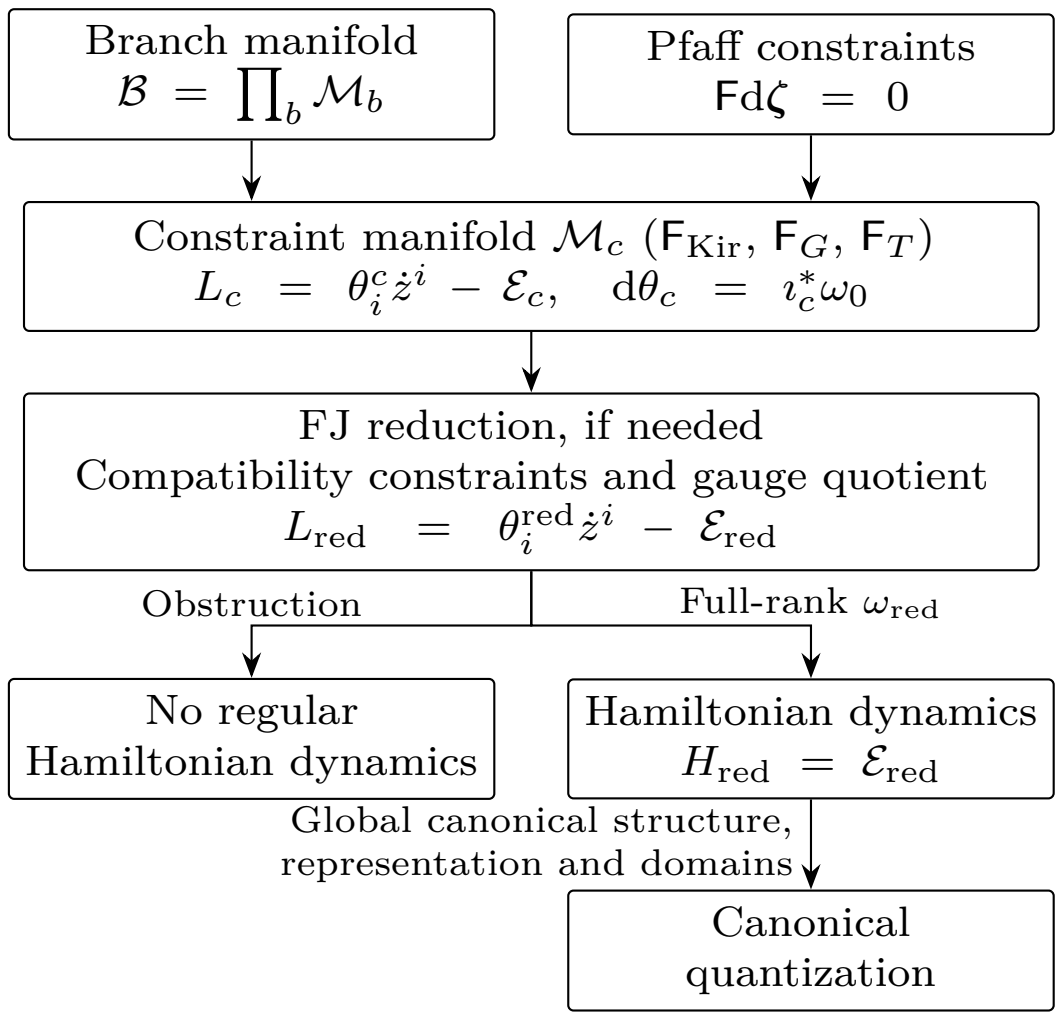
\begin{figure}[t]
\centering
\begin{tikzpicture}[font=\footnotesize,>=Stealth,
  box/.style={draw,rounded corners=1pt,align=center,inner sep=3pt}]
\node[box,text width=3.0cm] (B) at (-1.85,0)
  {Branch manifold\\$\mathcal B=\prod_b\mathcal M_b$};
\node[box,text width=3.0cm] (P) at (1.85,0)
  {Pfaff constraints\\$\mathsf F\dd\bm\zeta=0$};
\node[box,text width=6.5cm,anchor=north] (M)
  at ([xshift=1.85cm,yshift=-4mm]B.south)
  {{Constraint manifold $\mathcal M_c$ ($\mathsf{F}_{\rm Kir}$, $\mathsf{F}_{G}$, $\mathsf{F}_{T}$)}\\
   $L_c=\theta_i^c\dot{z}^i-\mathcal E_c,\quad \dd\theta_c=\imath_c^*\omega_0$ };
\node[box,text width=6.5cm,anchor=north] (R)
  at ([yshift=-4mm]M.south)
  {FJ reduction, if needed\\Compatibility constraints and gauge quotient\\
   $L_{\rm red}=\theta_i^{\rm red}\dot{z}^i-\mathcal E_{\rm red}$};
\node[box,text width=3.0cm,minimum height=0.7cm] (N)
  at ([xshift=-1.85cm,yshift=-1.7cm]R.center)
  {No regular\\Hamiltonian dynamics};
\node[box,text width=3.0cm,minimum height=0.7cm] (H)
  at ([xshift=1.85cm,yshift=-1.7cm]R.center)
  {Hamiltonian dynamics\\$H_{\rm red}=\mathcal E_{\rm red}$};
\node[box,text width=3.0cm] (Q) at ([yshift=-1.4cm]H.center)
  {Canonical quantization};
\draw[->] (B.south) -- (B.south |- M.north);
\draw[->] (P.south) -- (P.south |- M.north);
\draw[->] (M.south) -- (R.north);
\coordinate (branch) at ([yshift=-0.95cm]R.center);
\draw[->] (R.south) -- (branch) -| (N.north);
\draw[->] (R.south) -- (branch) -| (H.north);
\node[font=\scriptsize] at ([xshift=-1.85cm,yshift=-0.8cm]R.center)
  {Obstruction};
\node[font=\scriptsize] at ([xshift=1.85cm,yshift=-0.8cm]R.center)
  {Full-rank $\omega_{\rm red}$};
\draw[->] (H.south) -- node[midway,left,align=right,font=\scriptsize]
  {Global canonical structure,\\representation and domains} (Q.north);
\end{tikzpicture}
\caption{From branch data to Hamiltonian dynamics and quantization in Refs.~\cite{ParraRodriguez:2024,ParraRodriguez:2025}. Note that all linear constraints are integrated at the same (second) stage here, unlike in Ref.~\cite{Salcedo:2026}. The last step requires both global structure and operator data.}
\label{fig:topology}
\end{figure}

Reference~\cite{Salcedo:2026} (SCHO) presents a flux--charge Lagrangian, a
constraint taxonomy, and nonlinear gyrator substitutions as advances
over earlier quantization schemes. SCHO acknowledges the general scope of FJ quantization~\cite{ParraRodriguez:2025}, yet does not cite Ref.~\cite{ParraRodriguez:2024}, which already treats nonlinear nonreciprocal
circuits, and pure and mixed charge--flux null directions,
and incorporates compact
variables for reciprocal circuits. The formalism of Ref.~\cite{ParraRodriguez:2024} is
summarized in
Fig.~\ref{fig:topology}. Reference~\cite{ParraRodriguez:2025} extends
that construction to infinite-dimensional circuits. Transfer of nonlinear elements via the gyrator also appears in  Refs.~\cite{Rymarz:2022,Rymarz:2023PhD}.

\paragraph*{On context and interpretation.} The presentation of the context in the introduction of
SCHO conflates distinct issues. Their
Ref.~24~(here~\cite{Egusquiza:2025}) challenges the parasitic limit
rather than proposing regularization. Limitations of the 2019
construction~\cite{ParraRodriguez:2019} do not apply to the later
Faddeev--Jackiw (FJ)
method~\cite{ParraRodriguez:2024,ParraRodriguez:2025}. The presentation also makes distinctions that are unfounded, such as Lagrangian versus geometric perspectives. Lagrangian mechanics is a geometric theory \cite{arnol2013mathematical}. Furthermore, the FJ  method~\cite{ParraRodriguez:2024,ParraRodriguez:2025} uses first-order Lagrangians as the starting point for FJ degeneracy reduction, mostly to emphasize the local realization of the geometrical objects, even before reduction, directly from the energy function and the (pre-symplectic) two-form, and constructs the pertinent Lagrangians at all steps of the reduction
process. The circuit's structural information is encoded in the initial branch two-form and the Kirchhoff and other constraints, and is incorporated into the reduced two-forms and Lagrangians as these constraints are imposed. The difference with the reciprocal flux--charge formulation~\cite{Osborne:2024b} (their Ref.~26) and its Lagrangian is that the Kirchhoff reduction is carried out implicitly, by choosing as the starting point node fluxes and, typically, mesh charges (loop charges 
associated with faces of a planar embedding) and giving a recipe for the Lagrangian in those variables for the class of
circuits they consider (with planar connected components in SCHO). The recipe is precisely the result of the FJ method up to Kirchhoff reduction for that class of circuits, and also recovers the earlier reciprocal construction of those authors~\cite{Osborne:2024a} (Ref.~29 in~SCHO).  No new reduction or quantization principle is required. The local flux--charge formulation furthermore does not by itself specify the inherited global (topological) identifications, which flow through Kirchhoff reduction in~\cite{ParraRodriguez:2024,ParraRodriguez:2025}.

\begin{table*}[t]
\small
\centering
\setlength{\tabcolsep}{4pt}
\begin{tabular}{p{0.30\textwidth}p{0.7\textwidth}}
\hline
Reference~\cite{Salcedo:2026} & Relation to established results\\
\hline
\multicolumn{2}{l}{\textbf{Established theory and method}}\\[3pt]
First-order dynamics and Hamiltonian, Eqs.~(1.15)--(1.16), (2.30)--(2.31)
& First-order Lagrangians, FJ reduction, and Darboux coordinates: Ref.~\cite{ParraRodriguez:2024}, Eq.~(11), and Ref.~\cite{ParraRodriguez:2025}, Eq.~(6)\\[3pt]
Gyrator term, Eqs.~(2.1)--(2.3)
& Second-order treatment: Ref.~\cite{Solgun:2015b}, Sec.~6.4, and Refs.~\cite{ParraRodriguez:2019,ParraRodriguez:2022}. First-order constraint pullback: Appendix~\ref{app:summ-prev-work}~\cite{ParraRodriguez:2024,ParraRodriguez:2025}\\[3pt]
Constraint classification, Sec.~2.2
& $W(H)=0$ for $W\in\ker\omega$: Ref.~\cite{ParraRodriguez:2024}, Sec.~4, and Ref.~\cite{ParraRodriguez:2025}, Eq.~(7)\\[3pt]
Duality and nonlinear transfer, Secs.~3.1--3.2
& Pullback of the two-form and energy~\cite{ParraRodriguez:2024,ParraRodriguez:2025}. Explicit JJ flux-to-charge-cosine transfer: Refs.~\cite{Rymarz:2022,Rymarz:2023PhD}\\[3pt]
Reduced topology, Sec.~2.2
& Branch topology and constraint integration: Ref.~\cite{ParraRodriguez:2024}, Sec.~5, and Ref.~\cite{ParraRodriguez:2025}, Sec.~II C\\[3pt]
\hline
\multicolumn{2}{l}{\textbf{What Ref.~\cite{Salcedo:2026} adds}}\\[3pt]
Particular reciprocal realizations, Theorems~6--9
& Graph recipes composing established dualities~\cite{Rymarz:2022,Egusquiza:2022,ParraRodriguez:2024}, including nonlinear energies and, for bipartite graphs, ideal transformers~\cite{Belevitch:1968} (Appendix~\ref{app:theorems})\\[3pt]
Python code, \texttt{fluxcharge}~\cite{Salcedo:2026b}
& Automates established reduction.\\
\hline
\end{tabular}
\caption{Established construction and the applications added in SCHO. The local comparison is made on the same integral manifold of the constraints. The graph recipes are not restricted to quadratic energies.}
\label{tab:correspondence}
\end{table*}

\paragraph*{On gyrators, and reduction.} The introduction also identifies reciprocal realizations for some
circuits involving gyrators. Passive network
synthesis~\cite{Foster:1924,Cauer:1926,Brune:1931,Tellegen:1948,Newcomb:1966,Anderson:1968}
already provides minimal reactive realizations of lossless linear multiport responses with ideal transformer
and gyrator constraints. These results simplify quantization of linear
environments coupled to nonlinear
ports~\cite{Nigg:2012,Solgun:2014,Solgun:2015,ParraRodriguez:2019,Labarca:2024,Gigon:2026}.

These topics and others are expanded on in the main text of~SCHO. 
We summarize in 
Table~\ref{tab:correspondence} those claims, and put them in correspondence with  established methods. We use SCHO v2 PDF numbering to reference
equations and sections.

The main issue with respect to novelty, however, is the following:
\emph{if} the gyrator constraints are imposed, then the gyrator
Lagrangian term presented in Eqs. (2.1) and (2.3)
of~SCHO is irrelevant for the local equations, because it is locally a total derivative, as shown in Appendix~\ref{app:summ-prev-work}, and the construction is subsumed in the
formalism of~\cite{ParraRodriguez:2024}. The contention of~SCHO is that the gyrator term being included results in the gyrator constraints being part of the equations of motion, with
a caveat. If that is indeed the case, then again any further constraint reduction is also \emph{subsumed} in the FJ formalism~\cite{ParraRodriguez:2024,ParraRodriguez:2025}.

The caveat presented in the text is actually incomplete.
A deeper analysis (intimated by~SCHO) for ideal LCG circuits with constant gyrators shows, first, that if
an Euler--Lagrange solution exists that violates a gyrator constraint, then the associated two-form is necessarily degenerate and possesses a gyrator-mixing null vector that, second, is gauge in the terminology of~\cite{ParraRodriguez:2024}, in that its application to the energy function is identically zero. Therefore there exists a local coordinatization in which a coordinate along the null vector drops out of the
Lagrangian up to a total derivative. Full reduction for this circuit class applied to the two-form as extended by SCHO,  following the procedure of
Refs.~\cite{ParraRodriguez:2024,Osborne:2024a,Osborne:2024b}, quotients out this gauge
direction. The local dynamics are equivalent to those that follow from the FJ
formalism~\cite{ParraRodriguez:2024,ParraRodriguez:2025},
without anything new to show for it, for all gauge choices, as one of the choices is imposing the missing gyrator constraint, and all gauge choices are dynamically equivalent.

\paragraph*{On topology.}
When it comes to the topology of configuration
space,~SCHO states
that ``the Lagrangian formalism [...] acts as a mnemonic [...]
ultimately providing a [reduced space that can be outfitted with] a symplectic form'', thus
stressing the actual connection with the geometric perspective. 

They also point out that there is no intrinsic mathematical
argument to select the topology of configuration space. However, requiring the geometrical integration of the Kirchhoff constraints to recover the physical compactness of fluxes and  charges across circuits built from capacitors, inductors, Josephson junctions and phase-slip elements uniquely fixes the branch-topology assignment of Refs.~\cite{ParraRodriguez:2024,ParraRodriguez:2025}, within product branch manifolds with factors in $\{\mathbb{R}, S^1\}$. The corresponding discrete translational identifications then follow algebraically from the constraints, consistently with the presence or absence of superconducting islands~\cite{Devoret:2021}.

\paragraph*{Circuit families.}
A sizable part of SCHO involves the existence of
reciprocal circuits equivalent to those involving gyrators. Yet   non-reciprocity is primarily a property
of the responses of a system to external probing, and need not be
reflected in its spectrum. Therefore, in the present context, there
are reciprocal circuit realizations that are dynamically equivalent to
circuits with non-reciprocal elements. This is not exclusive to
circuits: in the Landau problem of a particle on the plane with a perpendicular constant and homogeneous magnetic field, which breaks time-reversal invariance, the dynamics is canonically equivalent to a harmonic  oscillator~\cite{ParraRodriguez:2022}, plus a (non-dynamical) conserved canonical pair.

For the circuits considered here, all energetic branches are
reciprocal. Nonreciprocal devices enter through constraints.
General regular reduction  results in canonical variables without making assumptions
about forests or bipartitions, as long as the circuit is well-posed,
thus allowing $QQ$, $\Phi\Phi$, and $Q\Phi$ quadratic couplings and
general nonlinear
energies~\cite{ParraRodriguez:2024,ParraRodriguez:2025}. However, canonical equivalence to a reciprocal circuit
does not imply reciprocal response at the original ports. Their
Definition~2 instead compares closed Hamiltonians without preserving
those port observables. More explicitly, Theorems~6--9 compose known
dualities. A forest has no closed chain of gyrator links between
components. A bipartite graph divides the components into two sets,
with every gyrator joining opposite sets. The recipes retain
transformed nonlinear energies and introduce Belevitch transformers in
the bipartite case~\cite{Belevitch:1968}. For positive-definite
quadratic Hamiltonians on $\R^{2n}$, Williamson reduction already
gives an LC oscillator realization without graph
restrictions~\cite{Williamson:1936,Egusquiza:2022}. Regular
second-order nonreciprocal quantization also predates
SCHO~\cite{Solgun:2015b,ParraRodriguez:2019}. Appendix~\ref{app:theorems}
separates these statements. The accompanying code~\cite{Salcedo:2026b} has been presented to automate the established reduction.

\paragraph*{Conclusions.}
SCHO adds graph recipes for particular reciprocal realizations and software. Its first-order formulation, constraint reduction, and nonlinear gyrator substitutions follow from the established construction~\cite{ParraRodriguez:2024,ParraRodriguez:2025}. These applications introduce no new reduction or quantization principle. The procedure put forward  by SCHO in fact adds complication to the process without any new result. Omitting Ref.~\cite{ParraRodriguez:2024} obscures the origin of the method.
Similar overlaps with Refs.~\cite{Egusquiza:2022,ParraRodriguez:2024} appear in Ref.~\cite{Smitham:2024}. We hope this comment helps to alleviate the issue.

\begin{acknowledgments}
A. P.-R. acknowledges support from the European Union's Marie Skłodowska-Curie Actions (MSCA) under grant agreement No.~101204967 (FTMcQED). I. L. E. acknowledges support from the Basque Government through Grant No.  IT1887-26. 
\end{acknowledgments}

\paragraph*{Declaration of use of AI.}
The authors developed the ideas. GPT and Claude initially computed the examples using our published method and assisted with  writing. The authors verified those derivations and take responsibility for the manuscript.

\appendix

\section{Summary of previous work and comparisons}
\label{app:summ-prev-work}
We present a summary of previous work to provide context and to fix
some notation. We will not fully adhere to the notation of our previous
work nor can we completely follow that of~SCHO.

\paragraph*{Notation and general framework.}
The geometric formulation, beyond being intrinsic, is realized in coordinates. Thus there are algebraic objects that are local
expressions of global geometric ones, and the constant forms and linear  constraints used here admit a matrix description. Thus, for two-forms we write $\omega=\frac12\Omega_{ij}\dd z^i\wedge\dd z^j$ in terms of local coordinates,  summing repeated indices. Only the antisymmetric part of $\Omega$ contributes. Along the same lines, the geometric assertion $W\in\ker\omega$ has
the local algebraic expression  $\Omega W=0$, where we use $W$ both for the geometric vector, with local expression $W=
W^i(\partial/\partial z^i)$, and for the algebraic vector of
components, and $\Omega$ is antisymmetric. The emphasis on geometry in Ref.~\cite{ParraRodriguez:2024} flows from two sources. 

First, it makes the statement of some properties and proofs more compact, and secondly it accommodates  non-trivial topologies in the different kinematic and dynamic manifolds. As was made clear in Ref.~\cite{ParraRodriguez:2024}, the only
possible manifolds at play are of the form
$\mathbb{R}^{d-k}
\times\left(S^1\right)^k$, all the way
from  the initial branch manifold to the manifold
obtained after integrating all Kirchhoff constraints. For $B$ branches, $d=2B$ initially and $d=B$ after
this reduction, and $k$ can change through the process. This manifold structure, together with
that of the constraints themselves, allows us to conclude the
integrability of those constraints. For practical purposes this means
that the algebraic analysis, with matrices, is adequate, as long as
the compact directions are properly taken into account. Gyrator and transformer constraints are more delicate in the presence of compact directions, but are also integrable for topologically trivial cases.

Constraint restriction locally means the identification of a regular submanifold $\mathcal{M}_S$ of
an initial larger manifold $\mathcal{M}$, with embedding map 
$\imath$ (we shall also use $\jmath$, with or without subscripts).  The pullback, denoted as  $\imath^*$ ($\jmath^*$, etc.), maps forms on $\mathcal{M}$ to forms on $\mathcal{M}_S$. Gauge reduction additionally takes a quotient by the null directions.

A pair $(\omega,E)$ with $\omega$ a closed two-form and $E$ a function (energy function) determines a local Lagrangian up to a total derivative.  For the cases
at hand, the initial two-form, $\omega_0$, is closed, because the coordinate one-forms are closed (even if not exact when associated with a compact direction, in spite
of the (standard) notation as if they were exact) and the coefficient matrix is constant on the manifold. Because pullback commutes with the exterior derivative, the reduced two-forms are also closed. A closed two-form $\omega$
is locally exact, that is, for each sufficiently small coordinate patch it can be written as $\omega=\dd\theta$. Let $\theta=\theta_i\dd\zeta^i$. Then we associate a (local) first-order Lagrangian term $L_\theta=\theta_i\dot{\zeta}^i$, and the total Lagrangian is $L_\theta-E$. This construction of the Lagrangian term with derivative is applicable at each stage of the reduction.

We shall use the subscript $b$ to be generic for branch (port)
variables. Branches are classified (for our limited purposes here) as
charge dependent ($\mathcal{C}$, general subscript $c$), flux dependent
($\mathcal{L}$ or $\mathcal{I}$, general subscript $l$), transformer ($\mathcal{T}$) and
gyrator ($\mathcal{G}$, general subscript $g$).
A port is a terminal pair with voltage $\dot\phi_b$ and current $\dot q_b$. For charge-dependent ($\mathcal C$) and flux-dependent ($\mathcal L$) branch energies, Ref.~\cite{ParraRodriguez:2024}, Eq.~(5), assigns
\begin{equation}
 \omega_0=\frac12\left(\sum_{c\in\mathcal C}\dd q_c\wedge\dd\phi_c
 +\sum_{l\in\mathcal L}\dd\phi_l\wedge\dd q_l\right).
 \label{eq:branch}
\end{equation}
Note that $\omega_0\equiv \omega_{2B}$ in Refs.~\cite{ParraRodriguez:2024,ParraRodriguez:2025}. Collecting branch variables as $\bm\zeta=(\bm q^T,\bm\phi^T)^T$, the
lossless linear constraints are $\mathsf F\dd\bm\zeta=0$, with
\begin{equation}
 \mathsf F=
 \begin{pmatrix}
  \mathsf F_{\rm Kir}\\
  \mathsf F_G\\
  \mathsf F_T
 \end{pmatrix}
 =
 \begin{pmatrix}
  \mathsf F_{\rm KCL}&0\\
  0&\mathsf F_{\rm KVL}\\
  \mathsf A_G&\mathsf B_G\\
  \mathsf T_q&0\\
  0&\mathsf T_\phi
 \end{pmatrix},
 \label{eq:jointconstraints}
\end{equation}
with Kir denoting Kirchhoff constraints, $G$ gyrators, and $T$
transformers. From standard circuit theory,
$\mathsf F_{\rm KCL}=\mathsf K_{\rm cut}^T$ and
$\mathsf F_{\rm KVL}=\mathsf K_{\rm loop}^T$, where
$\mathsf K_{\rm cut}$ and $\mathsf K_{\rm loop}$ span the cut and
cycle spaces. The transformer block separates current and voltage
constraints. Device blocks (gyrator and transformer blocks) are zero outside their port
branches. Taking the columns of $\mathsf K$ to span $\ker\mathsf F$
gives locally $\dd\bm\zeta=\mathsf K\dd\bm z$ and
$\Omega=\mathsf K^T\Omega_0\mathsf K$, with $\Omega_0$ representing
$\omega_0$.  This local
algebraic construction extends globally for the Kirchhoff constraints as proven
in~\cite{ParraRodriguez:2024,ParraRodriguez:2025},  by integrating the constraints and retaining the induced global identifications. Once an integral manifold is chosen, the
energy function is also restricted to that same
manifold~\cite{ParraRodriguez:2024,ParraRodriguez:2025}. In
~\cite{ParraRodriguez:2024,ParraRodriguez:2025} we wanted to emphasize
that all lossless linear constraints can be solved in one fell swoop,
and that the issue is the determination of the complete kernel, that
is the intersection of the separate kernels. The order in which this intersection
is determined is irrelevant. 

\paragraph*{Comparison with~\cite{Osborne:2024b}.}
SCHO, Sec.~1.2, instead (implicitly) first solves Kirchhoff's laws with node fluxes and mesh charges, then reduces according to the limited set of device constraints they consider. This staged procedure will recover the same local kernel and is novel only insofar as the selection of one particular order is novel. Furthermore, were there assumptions on the topology of the branch variables, it
would not, by itself, identify the topology of the Kirchhoff reduced manifold, which our construction predicts algorithmically.

The new element in~SCHO with respect to
\cite{Osborne:2024b} is the consideration of gyrators. Notice, by the
way, that in the formalism
of~\cite{ParraRodriguez:2024,ParraRodriguez:2025} the word gyrator was
used in lieu of general ideal nonreciprocal constraints, not restricted
to the four-terminal, two-port ideal gyrators, whereas that is the
only type considered in~SCHO. In the flux-charge
formalism \cite{Osborne:2024b}, Salcedo et al. claim that the introduction of these gyrators
necessitates the prescription of additional terms in the Lagrangian,
Eq.~(2.1) of~SCHO.

In order to compare notations and results, we relate the
starting point of~\cite{Osborne:2024b}, the use of node fluxes and
loop charges together with the term $\left\langle q\mid M\mid
  \dot{\phi}\right\rangle$, with $\jmath_{\rm Kir}^*(\omega_0)$, where $\jmath_{\rm Kir}$ embeds the
Kirchhoff constraint manifold in branch space.

In explicit local terms,
using  loop charges $\mathbf Q$ and node fluxes $\boldsymbol\Phi$,
the Kirchhoff constraints are solved by writing
\begin{equation}
 \dd\boldsymbol{\zeta}=
 \begin{pmatrix}
   \mathsf{F}_{\rm KVL}^T&0\\0&\mathsf{F}_{\rm KCL}^T
 \end{pmatrix}
 \begin{pmatrix}
   \dd\mathbf{Q}\\ \dd\boldsymbol{\Phi}
 \end{pmatrix}\,.
\end{equation}
The fundamental cut-set matrix $\mathsf{F}_{\rm KCL}$ and transposed incidence matrix $A^T$ have the same row space, and the fundamental loop matrix $\mathsf{F}_{\rm KVL}$ has the same row space as $B$. Here we consider $A$ and $B^T$ to be  full column rank, by setting aside one reference node flux and one outer mesh charge per connected component. Therefore, choosing the variables accordingly and connecting with the notation in SCHO,
\begin{equation}
 \dd\boldsymbol{\zeta}=
 \begin{pmatrix}
   B^T&0\\0&A
 \end{pmatrix}
 \begin{pmatrix}
   \dd\mathbf{Q}\\ \dd\boldsymbol{\Phi}
 \end{pmatrix}\,.
\end{equation}
References~\cite{Osborne:2024b,Salcedo:2026} use lower-case
letters for these variables.
The two-form $\omega_0$ is
\begin{equation}
  \omega_0= \frac{1}{2}\dd\boldsymbol{\zeta}^T\wedge
  \begin{pmatrix}
    0& \mathbb{P}_C\\ \mathbb{P}_I&0
  \end{pmatrix}\dd\boldsymbol{\zeta}\,,
\end{equation}
so the implementation of the Kirchhoff constraints reads
\begin{align}
\begin{aligned}
    \jmath_{\rm Kir}^*(\omega_0)&=
    \frac{1}{2}\dd\mathbf{Q}^T\wedge B(\mathbb{P}_C-\mathbb{P}_I)A\dd\boldsymbol{\Phi}\\
  &= \dd\mathbf{Q}^T\wedge M\dd\boldsymbol{\Phi}\,.
\end{aligned}  
\end{align}
We have introduced some notation from SCHO, to reach
precisely Eq.~(1.15), since this two-form maps into the $\left\langle
  Q\mid M\mid\dot{\Phi}\right\rangle$ term of the Lagrangian.

\paragraph*{Comparison with section 2 of~SCHO.}\label{sec:comp-with-sect}
The new element of SCHO  is the addition of a gyrator kinetic term to $\mathbf{Q}^T M\dot{\boldsymbol{\Phi}}$, a new term that we now show is superfluous.
For the gyrator constraints, we use the ordered-port convention
of SCHO, Eqs.~(1.14a)--(1.14b). Using branch variables, and the index $\gamma$ to identify each gyrator,
with $\bm i_\gamma=\dot{\bm q}_\gamma$ and
$\bm v_\gamma=\dot{\bm\phi}_\gamma$, the constitutive relations are
$\bm i_\gamma=\mathsf Y_\gamma\bm v_\gamma$ ($\bm v_\gamma=\mathsf Z_\gamma\bm i_\gamma$), with admittance and impedance
\begin{equation}
 \begin{aligned}
 \mathsf Y_\gamma
 &=-G_\gamma i\sigma_\gamma^y
   =G_\gamma\begin{pmatrix}0&-1\\1&0\end{pmatrix}=\mathsf Z_\gamma^{-1}.
 \end{aligned}
 \label{eq:gyrator-immittances}
\end{equation}
For the same port ordering and reference directions, this corresponds to a sign change in the usual convention (i.e., 
$v_2=R_\gamma i_1$, $v_1=-R_\gamma i_2$).

Explicitly, their Eq.~(2.3) introduces
\begin{align}
  L_g=\frac{1}{4}\sum_\gamma \left[G_\gamma
  \boldsymbol{\Phi}^TA^T\Gamma_\gamma
  A\dot{\boldsymbol{\Phi}}-
  G_\gamma^{-1}\mathbf{Q}^TB\Gamma_\gamma B^T\dot{\mathbf{Q}}\right]\,,
\end{align}
where $\gamma$ runs over all the gyrators and
$\Gamma_\gamma=\mathbb{P}_\gamma i\sigma_\gamma^y\mathbb{P}_\gamma$,
understanding the Pauli matrix as acting only on the two branch spaces
of the gyrator, with $\mathbb{P}_\gamma$ the projector on the gyrator
branch space. This $L_g$ corresponds to a one-form $\theta_g$, whose
exterior derivative $\omega_g$ is computed to be
\begin{equation}\label{eq:omegaggen}
\begin{aligned}
  \omega_g={}&\frac14\sum_\gamma\Bigl[
  G_\gamma\dd\boldsymbol{\Phi}^T\wedge A^T\Gamma_\gamma A\dd\boldsymbol{\Phi}\\
  &\quad-G_\gamma^{-1}\dd\mathbf Q^T\wedge B\Gamma_\gamma B^T\dd\mathbf Q\Bigr].
\end{aligned}
\end{equation}
The gyrator constraints (using SCHO's  convention) are
\begin{align}
  \mathbb{P}_\gamma\dd\mathbf{q}= -G_\gamma i\sigma^y_\gamma\mathbb{P}_\gamma\dd\boldsymbol{\phi}\,,
\end{align}
on branch variables, 
and imposing the Kirchhoff constraints on these equations results in
\begin{align}
  \mathbb{P}_\gamma B^T\dd\mathbf{Q}=
  -G_\gamma i\sigma^y_\gamma\mathbb{P}_\gamma A\dd\boldsymbol{\Phi}\,,
\end{align}
the gyrator constraints on the node flux and loop charge variables.
Insert these in $\omega_g$, Eq.~(\ref{eq:omegaggen}), and conclude
that
\begin{align}
  \dd\left[\imath_g^*(\theta_g)\right]&=\imath_g^*(\dd\theta_g)= \imath_g^*(\omega_g)\equiv0\,,
\end{align}
where $\imath_g$ is the inclusion map arising from imposing the
gyrator constraints. Thus $\imath_g^*\theta_g$ is locally exact. In other words, $\theta_g$
becomes locally exact on imposing the gyrator constraints. Therefore, were we to ignore it and forge ahead from the Lagrangian in (1.15), on imposing the gyrator constraints on it we would obtain exactly the same dynamics as those described  from the Lagrangian in (1.15) plus $L_g$ and imposing the gyrator constraints.

\paragraph*{On sections 2.1 and 2.2.}
The result in the previous paragraph provides us with interpretative
tools to read subsections 2.1 and 2.2 of~SCHO. We work locally with ideal LCG circuits, constant gyrators and energy functions that depend only on the branch charges of capacitive elements $\bm q_C$ and branch fluxes of inductive elements $\bm\phi_I$, $E=E(\bm q_C,\bm\phi_I)$. To impose Kirchhoff laws the branch variables are expressed using independent node-flux and loop-charge coordinates.
Define
\begin{align}
  \omega_S=  \dd\mathbf{Q}^T\wedge M\dd\boldsymbol{\Phi}+ \omega_g\,.
\end{align}
The gyrator constraints are integrable, and therefore a change of
coordinates exists into coordinates of an integral submanifold,
$\mathbf z$, on the one hand, and independent combinations of the integrals of the constraint one-forms
\begin{equation}
 \dd\bm c_\gamma= \mathbb P_\gamma B^T\dd\bm Q+G_\gamma i\sigma_\gamma^y
 \mathbb P_\gamma A\dd\bm\Phi,
 \label{eq:corrected-offsets-first}
\end{equation}
on the other. For each $\gamma$, which indexes gyrators,
there are two one-forms (independent when $\omega_S$ is full rank). If $\omega_S$ is full rank, $\dot{\bm c}_\gamma=0$ are part of the equations of motion. A system of coordinates $(\mathbf{z},\mathbf{c})$ can be chosen such that $\omega_S=\omega_z+\omega_c$, and such that the energy function does not depend on $\mathbf{c}_\gamma$, $E= E(\mathbf{z})$. $\omega_z$ is the lift of $\imath^*_g\circ\jmath^*_{\mathrm{Kir}}(\omega_0)$. The dynamical content of the system,  described by variables $\mathbf{z}$, is therefore exactly as if the FJ procedure had been followed, with no advantage.

If $\omega_S$ is not full rank, the situation becomes more
involved, and is  the subject of Section~2.2. In particular,
if it is not true that all gyrator constraints are implied by the
Euler--Lagrange equations of Lagrangian~(2.5), then there necessarily exist gyrator-mixing null vectors of $\omega_S$. Additionally, those gyrator-mixing null vectors associated with missing gyrator constraints annihilate the energy function.

Given this result, the gyrator relations not already implied by the
equations of motion can be imposed together by an admissible gauge choice.
Quotienting these gauge modes (together with imposing any remaining
null-vector constraints $W(E)=0$), fixing the conserved
gyrator integration constants, and reducing any resulting null
directions of the restricted two-form therefore yields the
same local reduced symplectic manifold
and dynamics as imposing the gyrator constraints upfront in the FJ
protocol, wherever the reduction is regular.

A minimal example of this process is that of a gyrator with a capacitor and an inductor shunting its ports. 
Choose the loop charge  $Q_i$ and node flux $\Phi_i$ variables to coincide with  the branch charge and flux of the shunt at port $i$,
with $C$ at port 1 and $L$ at port 2. Orient each shunt and its port alike. Then, following the formalism of SCHO as presented here,
\[
\begin{aligned}
\omega_S&=\frac{1}{2G}\,\dd(Q_1-G\Phi_2)\wedge\dd(G\Phi_1-Q_2),\\
E&=\frac{Q_1^2}{2C}+\frac{\Phi_2^2}{2L}.
\end{aligned}
\]
$\omega_S$ has rank two and is thus degenerate. This system is equivalent to two parallel capacitors. It possesses one energetic null
vector $W_-=G\partial_{Q_1}+\partial_{\Phi_2}$ and one pure gauge
mode $W_+=\partial_{\Phi_1}+G\partial_{Q_2}$, i.e., $W_+(E)\equiv0$.
This will impact the solutions by the appearance of
an arbitrary
function that does not enter the energy.  Indeed, 
the general solution of the Euler--Lagrange equations is 
$(Q_1,\Phi_2,\Phi_1,Q_2)=(a,-GLa/C,f(t),d+G f(t)-2Gat/C)$, where $a$ and $d$ are arbitrary constants and $f(t)$ is an arbitrary function. One of the gyrator constraints, $G\dot{\Phi}_2=\dot{Q}_1$, is part of the equations of motion, and is satisfied by all solutions. On the other hand, the second gyrator constraint is not zero, but rather $G\dot{\Phi}_1+\dot{Q}_2=2G\left[\dot{f}(t)-a/C\right]$. The choice $f(t)= a t/C$ guarantees that the second gyrator constraint is satisfied, and is equivalent dynamically to all other gauge choices. 
Quotienting $W_+$ and imposing $W_-(E)=0$ leaves,
for every gauge choice, the gauge-invariant 2D free-particle pair
$P=Q_1-G\Phi_2$ and $X=(G\Phi_1-Q_2)/(2G)$ with
$\omega_{\rm red}=\dd P\wedge\dd X$ and
$H_{\rm red}=P^2/[2(C+G^2L)]$. In the FJ protocol~\cite{ParraRodriguez:2024,ParraRodriguez:2025}, integrating the
gyrator constraints upfront selects fixed constants
$c_1=Q_1-G\Phi_2$ and $c_2=Q_2+G\Phi_1$, on which
$\imath_c^*\omega_0\equiv 0$ and reduction leaves a 0D 
sector. Fixing the conserved momentum $P=c_1=a(1+G^2L/C)$ in~SCHO
(with $c_2=d$ fixed by the gauge choice) collapses
$\omega_{\rm red}$ to zero and turns $\partial_X$ into a gauge
direction, establishing full equivalence of the reduced manifolds.

\paragraph*{General nonreciprocal devices.}
A more minor criticism towards~SCHO is that it only
considers two-port nonreciprocal ideal elements as primitive. This is completely correct, given Carlin's synthesis theorem~\cite{Carlin:1955}, yet would require finding the equivalent circuit when the original one includes more general non-reciprocal elements. On the other hand,
the earlier method incorporates directly general instantaneous linear lossless $N$-port constraints in the constant scattering form~\cite{ParraRodriguez:2024,ParraRodriguez:2025}
\begin{equation}
 R(\mathsf1+\mathsf S)\dd\bm q=(\mathsf1-\mathsf S)\dd\bm\phi.
 \label{eq:scattering}
\end{equation}
Here $R>0$ is a reference resistance and $\mathsf S$ is real and unitary (orthogonal), $\mathsf S^T\mathsf S=\mathsf1$. Equivalently, $\mathsf D\dd\bm q+\mathsf E\dd\bm\phi=0$, with $\mathsf D=R(\mathsf1+\mathsf S)$ and $\mathsf E=\mathsf S-\mathsf1$. These real $N\times N$ matrices satisfy $\operatorname{rank}[\mathsf D\ \mathsf E]=N$ and $\mathsf D\mathsf E^T+\mathsf E\mathsf D^T=0$, ensuring independent constraints and zero power $\dot{\bm q}^{T}\dot{\bm\phi}=0$ for currents directed into the device. For the devices in Eq.~\eqref{eq:jointconstraints}, $(\mathsf D\ \mathsf E)$ and the stacked $\mathsf F_G,\mathsf F_T$ blocks have the same kernel on the same port variables.

If $\mathsf1+\mathsf S$ is invertible, $\dd\bm q=\mathsf Y\dd\bm\phi$ with $\mathsf Y=-\mathsf D^{-1}\mathsf E$ and $\mathsf Y^T=-\mathsf Y$. A $-1$ eigenvalue obstructs this flux-only chart. Mixed charge--flux variables can remain necessary after imposing the circuit constraints. Darboux coordinates need not be physical branch fluxes.

\section{Reciprocal circuit realizations}
\label{app:theorems}
Theorems~6--9 concern realizations using specified circuit elements, not the existence of a reduced Hamiltonian. Their Definition~2 compares closed Hamiltonians without preserving the original ports. For positive-definite quadratic $H$ on $\R^{2n}$, Williamson diagonalization gives independent oscillators, hence an LC realization, irrespective of graph structure~\cite{Williamson:1936,Egusquiza:2022}. In the standard Schr\"odinger representation with symmetric ordering, the symplectic map is implemented unitarily. The port observables transform as well, so their response can remain nonreciprocal. This argument does not cover arbitrary nonlinear energies or compact phase spaces.

Their graph vertices are connected components  of the circuit graph (their really-connected components), with one edge per gyrator. A forest excludes closed chains of these links, including self-loops and parallel-edge cycles. Internal circuit loops remain allowed. Theorem~6 is additive separation of a detached reciprocal component, whose Hamiltonian and spectrum remain. Theorem~7 transfers a component attached by a single gyrator using the established substitutions~\cite{Rymarz:2022,Rymarz:2023PhD,Egusquiza:2022,ParraRodriguez:2024,ParraRodriguez:2025}
\begin{equation}
 h_C(q)\mapsto h_C(G\Phi^*),\qquad
 h_L(\phi)\mapsto h_L(Q^*/G),
 \label{eq:substitute}
\end{equation}
with inherited orientations and offsets. These identities do not assume quadratic energies. The remaining open-terminated gyrator is eliminated. Theorem~8 repeats this step, since every nonempty finite forest has a vertex of degree at most one, leaving a reciprocal realization with the transformed energies. Failure of this deletion rule does not establish that nonreciprocal elements are necessary.

For Theorem~9, bipartite coloring places exactly one port of
each gyrator in each color. Dualizing the components of one
color therefore converts every gyrator into an ideal
transformer through the known flux--charge
duality~\cite{Belevitch:1968,Egusquiza:2022}, including for nonlinear loads. These sufficient recipes add no quantization principle.

\bibliography{bibliography}
\end{document}